\documentclass[useAMS,usenatbib]{mn2e}
\usepackage{graphicx,amssymb,amsmath,times}
\usepackage{hyperref}
\title{Revisiting the hardening of the cosmic-ray energy spectrum at TeV energies}
\author[Satyendra Thoudam and J\"{o}rg R. H\"{o}randel]{Satyendra Thoudam\thanks{E-mail: s.thoudam@astro.ru.nl} and J\"{o}rg R. H\"{o}randel\\
Department of Astrophysics, IMAPP, Radboud University Nijmegen, P.O. Box 9010, 6500 GL Nijmegen, The Netherlands}

\defcitealias{Thoudam2012a}{Paper I}
\begin{document}
\date{}
\pagerange{}
\maketitle
\label{firstpage}
\begin{abstract}
Measurements of cosmic rays by experiments such as ATIC, CREAM, and PAMELA indicate a hardening of the cosmic-ray energy spectrum at TeV energies. In our recent work \citep{Thoudam2012a}, we showed that the hardening can be due to the effect of nearby supernova remnants. We showed it for the case of protons and helium nuclei. In this paper, we present an improved and more detailed version of our previous work, and extend our study to heavier cosmic-ray species such as boron, carbon, oxygen, and iron nuclei. Unlike our previous study, the present work involves a detailed calculation of the background cosmic rays and follows a consistent treatment of cosmic-ray source parameters between the background and the nearby components. Moreover, we also present a detailed comparison of our results on the secondary-to-primary ratios, secondary spectra, and the diffuse gamma-ray spectrum with the results expected from other existing models, which can be checked by future measurements at high energies.
\end{abstract}

\section{Introduction}
Recent measurements of cosmic rays by the ATIC \citep{Panov2007}, CREAM \citep{Yoon2011}, and PAMELA \citep{Adriani2011} experiments have indicated the presence of hardening in the energy spectra of protons, helium and heavier nuclei at TeV energies. The observed hardening does not seem to be in good agreement with general theoretical predictions. Based on the simple linear theory of diffusive shock acceleration (DSA) of cosmic rays \citep{Krymskii1977, Bell1978}, and the nature of cosmic-ray transport in the Galaxy (see e.g., \citealp{Ginzburg1976}),  the cosmic-ray spectrum is expected to follow a single power-law at least up to $\sim 3$ PeV, the so-called ``cosmic-ray knee".

Different explanations for the spectral hardening have been proposed. Most of these explanations suggest either hardening in the cosmic-ray source spectrum or changes in the propagation properties in the Galaxy. Possible scenarios that can produce a hardened source spectrum have been suggested in \citealp{Biermann2010}, \citealp{Ohira2011}, \citealp{Yuan2011}, and \citealp{Ptuskin2013}, and models based on propagation effects have been discussed in \citealp{Ave2009}, \citealp{Tomassetti2012}, and \citealp{Blasi2012}. Various interpretations of the spectral hardening including the effect of local sources can be found in  \citealp{Vladimirov2012}.
 
It is generally believed that nearby sources can affect the properties of cosmic rays observed at the Earth, and they might also account for the observed spectral hardening. In \citealp{Erlykin2012}, it was suggested that the hardening can be due to a steep local component dominating up to $\sim 200$ GeV/n, and a flatter background dominating at higher energies. Another scenario is that the high-energy spectrum might be dominating by the local component and the low-energy region is dominated by the background. This scenario was proposed in our recent work \citep[hereafter referred to as \citetalias{Thoudam2012a}]{Thoudam2012a}, in which we considered nearby supernova remnants within $1$ kpc in the Solar vicinity. In addition to explaining the spectral hardening, we could also explain the observed hardening of the helium spectrum at relatively lower energy/nucleon with respect to the protons. Moreover, we showed that the hardening may not continue beyond few tens of TeV/nucleon.   

In \citetalias{Thoudam2012a}, we concentrated only on protons and helium nuclei. Moreover, we did not perform a detailed calculation for the background component. The background was obtained by fitting the measured low-energy data in the range of $(20-200)$ GeV/n. Considering that this is the energy region where both the contribution of the local sources and the effect of the Solar modulation are expected to be minimum, the background thus obtained seems to represent fairly well the averaged cosmic-ray background present in the Galaxy. However, the background should be consistent with the observed data down to lower energies say up to sub-GeV or $\sim 1$ GeV energies. This  consistency was not checked in \citetalias{Thoudam2012a}. In addition, if we assume a common cosmic-ray source throughout the Galaxy, then the source parameters for the background component such as the spectral index and the cosmic-ray injection power should also be consistent with those of the nearby sources. To check whether the spectral hardening can still be explained by maintaining such consistencies, a detailed calculation should be carried out for both components together. This is the main motivation of our present paper. A similar study was recently presented in \citealp{Bernard2012}. The main difference between their work and ours is on the treatment of the nearby sources. They assumed supernova remnants as instantaneous point-like sources injecting cosmic rays in an energy-independent manner, while we consider finite-sized remnants producing cosmic rays of different energies at different stages in their lifetime. For distant sources, the much simpler energy-independent point-source approximation may represent a valid approximation, but for nearby sources, it is more realistic to take into account their finite sizes and the nature of cosmic-ray injection in the Galaxy \citep{Thoudam2012b}.

In the present work, we will further extend our study to heavier cosmic-ray species namely boron, carbon, oxygen, and iron. We will also present a detailed comparison of our results on the secondary-to-primary ratios, secondary spectrum and the diffuse gamma-ray spectrum with those expected from other existing models. Our model is described in section 2. Then, our calculations for the cosmic-ray spectrum from a nearby source and for the background spectrum are presented in sections 3 and 4, respectively, and in section 5, the various interaction cross-sections and the  matter density that will be used for our study will be presented. In section 6, constraints on the contribution of the nearby sources imposed by the secondary cosmic-ray spectrum are discussed. In section 7, the results on the heavy nuclei are presented, and in section 8, the results on protons and helium nuclei are described. In section 9, we discuss and compare our results with the predictions of other models.

\section{The model}
Although there is no direct evidence yet that proves that supernova remnants are the major source of Galactic cosmic rays, they represent the most probable candidates both from the theoretical and the observational point of views. Theoretically, it has been established that the DSA process \citep{Krymskii1977, Bell1978} inside supernova remnants can produce a power-law spectrum of particles up to very high energies with a spectral index close to $2$. This value of spectral index agrees nicely with the values determined from radio observation of supernova remnants \citep{Green2009}. Moreover, if $\sim (10-20)\%$ of the total supernova kinetic energy is channeled into cosmic rays, supernova remnants can easily account for the total amount of cosmic-ray energy contained in the Galaxy. Observationally, the presence of high-energy particles inside supernova remnants is evident from detections of non-thermal X-rays and TeV $\gamma$-rays from a number of remnants \citep{Parizot2006, Aharonian2006, Aharonian2008}. The non-thermal X-rays are most likely synchrotron radiations produced by high-energy electrons in the presence of magnetic fields. The TeV $\gamma$-rays are produced either by inverse compton interactions of high energy electrons with ambient low energy photons or by the decay of neutral pions, which are produced by the interaction of hadronic cosmic rays, mainly protons with the surrounding matter. Irrespective of the nature of production, the detection of TeV $\gamma$-rays suggests the presence of charged particles with energies larger than few TeV inside supernova remnants.
 
Based on these observational evidences and our current theoretical understanding, we assume that supernova remnants are the main sources of cosmic rays in the Galaxy. We further assume that cosmic rays observed at the Earth consist of two components: a steady background which dominates most of the spectrum and a local component which is produced by nearby sources. The background is considered to be produced by distant sources which follow a uniform and continuous distribution in the Galactic disk, and the local component is assumed to be contributed by supernova remnants with distances within $1$ kpc from the Earth. A list of supernova remnants that will be considered for our study are given in Table 1.
\begin{table}
\centering
\caption{\label{SNR} List of supernova remnants with distances $<1$ kpc considered in our study.}
\begin{tabular}{c|c|c}
\hline
Name	&	Distance (kpc)		&	Age (yr)\\         
\hline
Geminga	&	$0.15$	&	$3.4\times 10^5$\\
Loop1		&	$0.17$	&	$2.0\times 10^5$\\
Vela			&	$0.30$	&	$1.1\times 10^4$\\
Monogem     	&	$0.30$	&	$8.6\times 10^4$	\\
G299.2-2.9	&	$0.50$	&	$5.0\times 10^3$	\\
Cygnus Loop	&	$0.54$	&	$1.0\times10^4$\\
G114.3+0.3  	&	$0.70$	&	$4.1\times 10^4$	\\
Vela Junior	&	$0.75$	&	$3.5\times 10^3$\\
S147		&	$0.80$	&	$4.6\times 10^3$\\
HB9		&	$0.80$	&	$6.6\times 10^3$\\
HB21		&	$0.80$	&	$1.9\times 10^4$\\
SN185		&	$0.95$	&	$1.8\times 10^3$	\\
\hline
\end{tabular}
\end{table}

The two cosmic-ray components are treated with different propagation models. The background component is treated in the framework of a steady state diffusion model, and the local component in a time dependent model. For the background component, we adopt the model described in \citealp{Thoudam2008} where the cosmic-ray diffusion region is taken to be a cylindrical disk with infinite radial boundary and finite vertical boundaries $\pm H$. For a typical value of the radial boundary which is $\sim 20$ kpc or more, our assumption of infinite radial boundary represents a good approximation for cosmic rays at the Earth. It is because for large radial boundary, cosmic-ray escape is dominated by escape through the vertical boundary and the effect of the radial boundary on the cosmic-ray flux becomes negligible. Regarding the size of $H$, different models adopt different values in the range of $\sim (2-10)$ kpc \citep{Strong2010}, and for the present study, we assume $H=5$ kpc. Furthermore, both the background sources and the interstellar matter are assumed to be distributed in the Galactic plane in an infinitely thin disk of radius $R$. And we take $R=20$ kpc for our study.
 
For the local component, we assume a diffusion region with infinite boundaries in all directions. This assumption is made,  considering that the spectrum of cosmic rays from nearby sources do not depend much either on the radial or the vertical boundaries because of their very short propagation time to the Earth relative to the escape times from the Galactic boundaries \citep{Thoudam2007}. Strictly speaking, the propagation does depend on the boundary because of the dependence of the  diffusion coefficient on the vertical boundary. But, once the diffusion coefficient is fixed, one can neglect the dependencies  on the diffusion boundaries and determine the cosmic-ray flux using infinite boundaries.   
        
\section {Cosmic rays from a nearby source}
The propagation of cosmic rays from a nearby source is governed by the diffusive transport equation,
\begin{equation}
\nabla\cdot(D\nabla N)+Q=\frac{\partial N}{\partial t}
\end{equation}
The first and the second terms on the left represents the diffusion and the source terms respectively, with $N(\textbf{r},E,t)$ representing the particle number density, $E$ the kinetic energy/nucleon, $\textbf{r}$ the distance from the center of the source and $t$ representing the time. The diffusion coefficient $D$ is assumed to be a function of the particle rigidity $\rho$ as $D(\rho)=D_0\beta(\rho/\rho_0)^\delta$, where $\delta$ is the diffusion index, $\beta=v/c$ with $v$ denoting the particle velocity and $c$ is the velocity of light. For a nuclei carrying charge $Ze$ and mass number $A$, the rigidity can be written as $\rho=APc/Ze$, where $P$ denotes the momentum/nucleon of the nuclei. Thus, for the same $P$ ($\approx E$ for all the energies of our interests here), all heavy nuclei with charge $Z>1$ diffuse relatively faster than the protons by a factor of $(A/Z)^\delta$. In Eq. (1), we neglect the effect of interactions with the interstellar matter as the time taken for cosmic rays to reach the Earth is expected to be much less than the nuclear interaction timescales. Also, we do not include effects which are important only at low energies such as ionization losses, convection by the Galactic wind, and a possible re-acceleration by the interstellar turbulence. With these approximations, we assume that Eq. (1) effectively describes the propagation of cosmic rays with energies above $1$ GeV/n. 

As mentioned before, an important feature of our model (as also in \citetalias{Thoudam2012a}) is the assumption of finite-sized  sources which inject cosmic rays in an energy-dependent manner. Our model is based on the current basic understandings of the DSA theory \citep{Krymskii1977, Bell1978} inside supernova remnants. In DSA theory, charged particles are accelerated each time they cross the supernova shock front. During the acceleration process, a small fraction of the particles are advected downstream of the shock and cannot be further accelerated, while a major fraction diffuse upstream which can be again taken over by the expanding shock and continue their acceleration. Thus, efficient acceleration can be achieved when particles are effectively confined near the shock in the upstream region. It is now generally accepted that the confinement can be provided by magnetic scattering due to the turbulence generated ahead of the shock. Particles can remain confined as long as their upstream diffusion length is much less than the shock radius, i.e., $l_d\ll R_s$. The diffusion length is related to the upstream diffusion coefficient $D_s$ and the shock velocity $u_s$ as $l_d=D_s/u_s$. Then, for $D_s$ that scales with energy, for instance $D_s\propto E$ in the Bohm diffusion limit, $l_d\propto E/u_s$. This shows that high-energy particles can escape at early stages during the evolution of the remnant while the shock is still strong. For the lower energy particles, they can escape only at later stages when the shock becomes weak. Based on this general understanding, and at the same time considering that the actual energy dependence of $D_s$ is not clearly understood because of many complicated processes involved during the acceleration process such as magnetic field amplification and back-reaction of the particles on the shock, the cosmic-ray escape time is parameterized in a simple form as (\citetalias{Thoudam2012a})
\begin{equation}
t_{esc}(\rho)=t_{sed}\left(\frac{\rho}{\rho_{m}}\right)^{-1/\alpha}
\end{equation}
where $\rho$ represents the particle rigidity, $t_{sed}$ represents the onset of the Sedov phase, $\rho_{m}$ the maximum rigidity, and $\alpha$ is a positive constant which is taken as a parameter in the present study. The maximum particle energy is assumed to scale with its charge as $ZU_{m}$, where $U_{m}$ denotes the maximum kinetic energy for protons. We take $U_{m}=1$ PeV \citep{Berezhko1996}, which corresponds to $\rho_{m}=1$ PV. Expressing Eq. (2) in energy/nucleon, we get
\begin{equation}
t_{esc}(E)=t_{sed}\left(\frac{AE}{Ze\rho_{m}}\right)^{-1/\alpha}
\end{equation}
Eq. (3) shows that for the same $E$, all heavy nuclei escape at relatively early times compared to the protons by a factor of $(A/Z)^{-1/\alpha}$. This early escape of heavier nuclei in our model was the key to explain the observed spectral hardening of helium at lower energy/nucleon with respect to the protons in \citetalias{Thoudam2012a}.

At some later stage when the supernova shock becomes too weak to accelerate particles, the turbulence level in the upstream region goes down and the remnant can no longer hold any particles. At this point, all low-energy particles which remained confined until this stage escape into the interstellar medium. We assume this to occur when the supernova age becomes $10^5$ yr. Then, the complete cosmic-ray escape time is taken to be
\begin{equation}
T_{esc}(E)= \mathrm{min}\left[t_{esc}(E), 10^5 \mathrm{yr}\right]
\end{equation}
Details about the nature of particle escape from supernova remnants can be found in the literature (\citealp{Ptuskin2005},  \citealp{Caprioli2009}, etc.). Having parameterized the cosmic-ray escape times, the corresponding escape radii can be calculated as,
\begin{equation}
R_{esc}(E)=2.5u_0\;t_{sed}\left[\left(\frac{T_{esc}}{t_{sed}}\right)^{0.4}-0.6\right]
\end{equation}
where $u_0$ is the initial velocity of the shock.

Assuming that the supernova remnant is spherically symmetric, the source term in Eq. (1) is taken as
\begin{equation}
Q(\textbf{r},E,t)=\frac{q(E)}{A_{esc}}\delta(t-T_{esc})\delta(r-R_{esc})
\end{equation}
where $A_{esc}(E)=4\pi R_{esc}^2$ denotes the area of the remnant corresponding to the escape time of cosmic rays of kinetic energy/nucleon $E$, $r$ represents the radial variable, and $q(E)=Aq(U)$ is the source spectrum with $q(U)$ given by,
\begin{equation}
q(U)=\kappa(U^2+2Um)^{-(\Gamma+1)/2}(U+m)
\end{equation}
where $U=AE$ is the total kinetic energy of the particle, $\Gamma$ is the source spectral index, $m$ denotes the rest mass energy  of the particle, and $\kappa$ is a constant which denotes the cosmic-ray injection efficiency. In Eq. (7), we assume that particles are produced with a power-law momentum spectrum. It should be noted that non-linear DSA theory predicts a slight deviation from pure power-laws with a somewhat flatter spectrum at high energies \citep{Berezhko1994, Ptuskin2010}. In \citealp{Ptuskin2013}, this non-linear effect has been considered to explain the observed cosmic-ray spectral hardening at TeV energies. In the present model, we neglect such non-linear effects, and assume a pure power-law source spectrum.

With all the above ingredients, the solution of Eq. (1) is obtained as,
\begin{eqnarray}
N(r_s,E,t)=\frac{q(E)\,R_{esc}}{r_sA_{esc}\sqrt{\pi D(t-T_{esc})}}\mathrm{exp}\left[-\frac{\left(R_{esc}^2+r_s^2\right)}{4D(t-T_{esc})}\right]\nonumber\\
\times\mathrm{sinh}\left(\frac{r_sR_{esc}}{2D(t-T_{esc})}\right)
\end{eqnarray}
Eq. (8) gives the cosmic-ray spectrum at a distance $r_s$ (measured from the center) from a supernova remnant injecting cosmic rays of different energies at different times during its evolution. Using the properties $e^x\rightarrow 1$ and $\sinh(x)\approx x$ for very small $x$, it can be checked that at high energies where the diffusion radius $r_{diff}=\sqrt{D(t-T_{esc})}$ is much larger than $(r_s, R_{esc})$, Eq. (8) yields $N(E)\propto q(E)/D^{3/2}\propto E^{-\left(\Gamma+3\delta/2\right)}$.

\section{Cosmic-ray background spectrum}
\subsection{Primary cosmic rays}
The background component for a primary cosmic-ray species, hereafter denoted by $p$, can be calculated from the steady state diffusion-loss equation:
\begin{equation}
\nabla\cdot(D_p\nabla N_p)-\eta v_p\sigma_p\delta(z)N_p=-Q_p
\end{equation}
The terms on the left represent diffusion and losses due to inelastic collisions, where $\eta$ is the averaged surface density of interstellar matter on the Galactic disk, $v_p$ is the particle velocity and $\sigma_p(E)$ is the collision cross-section. We consider the diffusion region as a cylindrical disk bounded in the vertical directions at $z=\pm H$ and unbounded in the radial direction, as mentioned in section 2. For sources uniformly distributed in the thin Galactic disk, the source term is represented by a delta function as $Q_p(\textbf{r},E)=S q(E)\delta(z)$, where $S$ denotes the surface density of supernova explosion rate in the Galactic disk. Eq. (9) can be solved analytically as described in \citep{Thoudam2008}, and the solution at $r=0$ is given by
\begin{eqnarray}
N_p(z,E)=\frac{RS q(E)}{2D_p}F_p
\end{eqnarray}
where,
\begin{eqnarray}
F_p=\int^\infty_0\frac{\mathrm{sinh}[K(H-z)]}{\mathrm{sinh}(KH)\left[K\mathrm{coth}(KH)+\frac{\eta v_p\sigma_p}{2D_p}\right]}\times\mathrm{J_1}(KR)dK\nonumber
\end{eqnarray}
In Eq. (10), $\mathrm{J_1}$ denotes the Bessel function of order 1, and $R$ is the radial size of the supernova remnant distribution. Taking $z=0$, Eq. (10) gives the cosmic-ray spectrum at the Earth. This is reasonable considering the fact that the position of the Earth ($\sim 8.5$ kpc from the Galactic center) is well contained within the size of the source distribution  which is taken to be $20$ kpc, and also that the majority of the cosmic rays that reach the Earth are produced by sources within $\sim 5$ kpc, the size of the vertical halo boundary. It is important to mention that in Eq. (10), $q(E)$ is taken to be the same as in the case of the local component given in the previous section, thus maintaining the same source spectrum between the local and the background components.

\subsection{Secondary cosmic rays}
Cosmic-ray secondaries are produced as spallation products from the interaction of heavier primaries with the interstellar matter during their propagation through the Galaxy. For matter distribution on the thin Galactic disk, the secondary production rate can be calculated as,
\begin{equation}
Q_s(\textbf{r},E)=\int^{\infty}_{E}\eta v_pN_p(\textbf{r},E^\prime)\delta(z)\frac{d}{dE^\prime}\sigma_{ps}(E,E^\prime) dE^\prime
\end{equation}
where $s$ denotes secondary species, $N_p$ represents the primary number density, and $d\sigma_{ps}(E,E^\prime)/dE^\prime$ represents the differential cross-section for the production of a secondary nucleus with an energy/nucleon $E$ from the spallation of a primary nucleus with energy/nucleon $E^\prime$. Assuming that the energy/nucleon is conserved during the spallation process, the differential cross-section can be written as, 
\begin{equation}
\frac{d}{dE^\prime}\sigma_{ps}(E,E^\prime)=\sigma_{ps}\delta(E^\prime-E),
\end{equation}
where $\sigma_{ps}$ is the total spallation cross-section of the primary to the secondary. Eq. (11) then  reduces to
\begin{equation}
Q_s(\textbf{r},E)=\eta v_p\sigma_{ps}N_p(\textbf{r},E) \delta(z)
\end{equation}

The propagation of secondaries follows a similar equation that describes their primaries as given by Eq. (9), with the source term replaced by Eq. (13). Their differential number density is given by \citep{Thoudam2008}
\begin{eqnarray}
N_s(z,E)=\eta v_p\sigma_{ps}N_p(0,E)\frac{R}{2D_s}F_s
\end{eqnarray}
where $N_p(0,E)$ is given by Eq. (10), and
\begin{eqnarray}
F_s=\int^\infty_0\frac{\mathrm{sinh}[K(H-z)]}{\mathrm{sinh}(KH)\left[K\mathrm{coth}(KH)+\frac{\eta v_s\sigma_s}{2D_s}\right]}\times \mathrm{J_1}(KR)dK\nonumber
\end{eqnarray}
By taking $z=0$, Eq. (14) gives the background spectrum of secondary cosmic rays at the Earth. From Eq. (14), it can be shown that the ratio of the secondary to the primary densities for the background component gives a good measure of the cosmic-ray diffusion coefficient as,
\begin{equation}
\frac{N_s}{N_p}\propto \frac{1}{D_s}
\end{equation}
which is a well-known result in cosmic-ray propagation studies.

\section{Interaction cross-sections and interstellar matter density}
\subsection{Cross-sections}
In this section, we present the various interaction cross-sections and the interstellar matter distribution that will be used in our calculations. For protons, the inelastic interaction cross-section is taken from the simple parameterization given in \citealp{Kelner2006}:
\begin{equation}
\sigma_{P}(T)=\left(34.3+1.88L+0.25L^2\right)\left[1-\left(\frac{T_{th}}{T}\right)^4\right]^2 \mathrm{mb}
\end{equation}
where $T$ is the total energy of the cosmic-ray proton, $L=\ln (T/1\;\mathrm{TeV})$ and $T_{th}=1.22$ GeV is the threshold energy for the production of $\pi^0$ mesons. For helium and other heavier nuclei, the spallation cross-sections are taken from \citealp{Letaw1983}:
\begin{equation}
\sigma_A(E)=\sigma_0\left[1-0.62 e^{-E/0.2}\sin\left(10.9 X^{-0.28}\right)\right]
\end{equation}
where $A$ represents the mass number of the nuclei, $E$ is the kinetic energy/nucleon in GeV/n, $X=E/0.001$ and 
\begin{equation}
\sigma_0=45A^{0.7}\left[1+0.016\sin\left(5.3-2.63 \ln A\right)\right] \mathrm{mb}
\end{equation}
Both Eqs. (16) and (17) represent a good approximation to the measured cross-section data down to sub-GeV energies. For helium, as suggested by \citealp{Letaw1983}, we further include a correction factor of $0.8$ in Eq. (17).

For the secondary boron production, we consider only the $^{12}$C and $^{16}$O primaries as they dominate the total boron production in the Galaxy. Through spallation, they produce ($^{11}$B,$^{10}$B) and ($^{11}$C,$^{10}$C) isotopes. The latter further decay into ($^{11}$B,$^{10}$B) thereby contributing to the production of boron. For our calculations, we use the tabulated production cross-sections of these isotopes given in \citealp{Heinbach1995}.

\subsection{Matter density}
Since we assume that the interstellar matter is distributed in the thin Galactic disk, it is more relevant for our study to determine the surface matter density on the disk rather than the actual number density. For a given distribution of atomic hydrogen in the Galaxy $n_{HI}(r,z)$, the surface density at the Galacto-centric radius $r$ can be obtained as $n_{HI}(r)=\int_{-\infty}^{\infty} n_{HI}(r,z)dz$. Similarly, for the molecular hydrogen distribution, we can calculate the surface density as $n_{H_2}(r)=\int^{\infty}_{-\infty}n_{H_2}(r,z)dz$. The total surface density of atomic hydrogen is then obtained as $n_H(r)=n_{HI}(r)+2n_{H_2}(r)$. Because cosmic rays arriving at the Earth are mostly produced within a distance equivalent to the halo height $H=5$ kpc, only the interstellar matter distributed within a circle of radius $5$ kpc is important for our study. For our calculations, we use the averaged surface density determined for this circle.

The distribution of atomic hydrogen is taken from \citealp{Gordon1976} and \citealp{Cox1986}, while that of the molecular hydrogen is taken from \citealp{Bronfman1988}. From these distributions, we obtain the averaged surface density of atomic hydrogen within our $5$ kpc circle as $2.85\times 10^{20}$ atoms cm$^{-2}$, and that of molecular hydrogen as $1.16\times 10^{20}$ molecules cm$^{-2}$. This gives a total averaged surface density of atomic hydrogen of $5.17\times 10^{20}$ atoms cm$^{-2}$ which is finally used for the present study. In addition, it is assumed that the interstellar matter consists of $10\%$  helium.

\section{Constraint on the contribution of nearby sources}
Before we proceed, we will first determine the diffusion coefficient of cosmic rays in the Galaxy. As already mentioned in section 4.2, the diffusion coefficient $D(\rho)=D_0\beta(\rho/\rho_0)^\delta$ can be determined from the secondary-to-primary ratio. For our study, we choose the boron-to-carbon ratio since this is the most well-measured and well-studied ratio. The boron-to-carbon ratio calculated using Eq. (14) is compared with the measured data in Figure 1. The solid line represents our calculation, and the data are taken from HEAO \citep{Engelmann1990}, CRN \citep{Swordy1990}, CREAM \citep{Ahn2008} and TRACER \citep{Obermeier2011}. We find that choosing values of $D_0=1.55\times 10^{28}$ cm$^2$ s$^{-1}$, $\rho_0=3$ GV and $\delta=0.54$ produces a good fit to the data. Our calculation takes into account the effect of solar modulation using the force field approximation with modulation parameter $\phi=400$ MV \citep{Gleeson1968}.

Both the values of $D_0$ and $\delta$ obtained in this work are lower than the values adopted in \citetalias{Thoudam2012a} for the case of pure diffusion model\footnote{For the rest of this paper, unless otherwise stated, any comparison with \citetalias{Thoudam2012a} will be always with the pure diffusion model (Model A of \citetalias{Thoudam2012a}). Comparison for the re-acceleration model (Model B of Paper I) will not be presented here as the comparison will look similar to what we obtain in the case of  the pure diffusion model.}. We used $D_0=2.9\times 10^{28}$ cm$^2$ s$^{-1}$ and $\delta=0.6$ in \citetalias{Thoudam2012a}. These values were taken from \citealp{Thoudam2008} which were determined using a slightly larger value of interstellar hydrogen density, and based on earlier measurements before CREAM and TRACER data became available. In Figure 1, it can be noticed that our new value of $D$ nicely agrees with the  measurements up to $\sim (1-2)$ TeV/n. 
\begin{figure}
\centering
\includegraphics*[width=0.315\textwidth,angle=-90,clip]{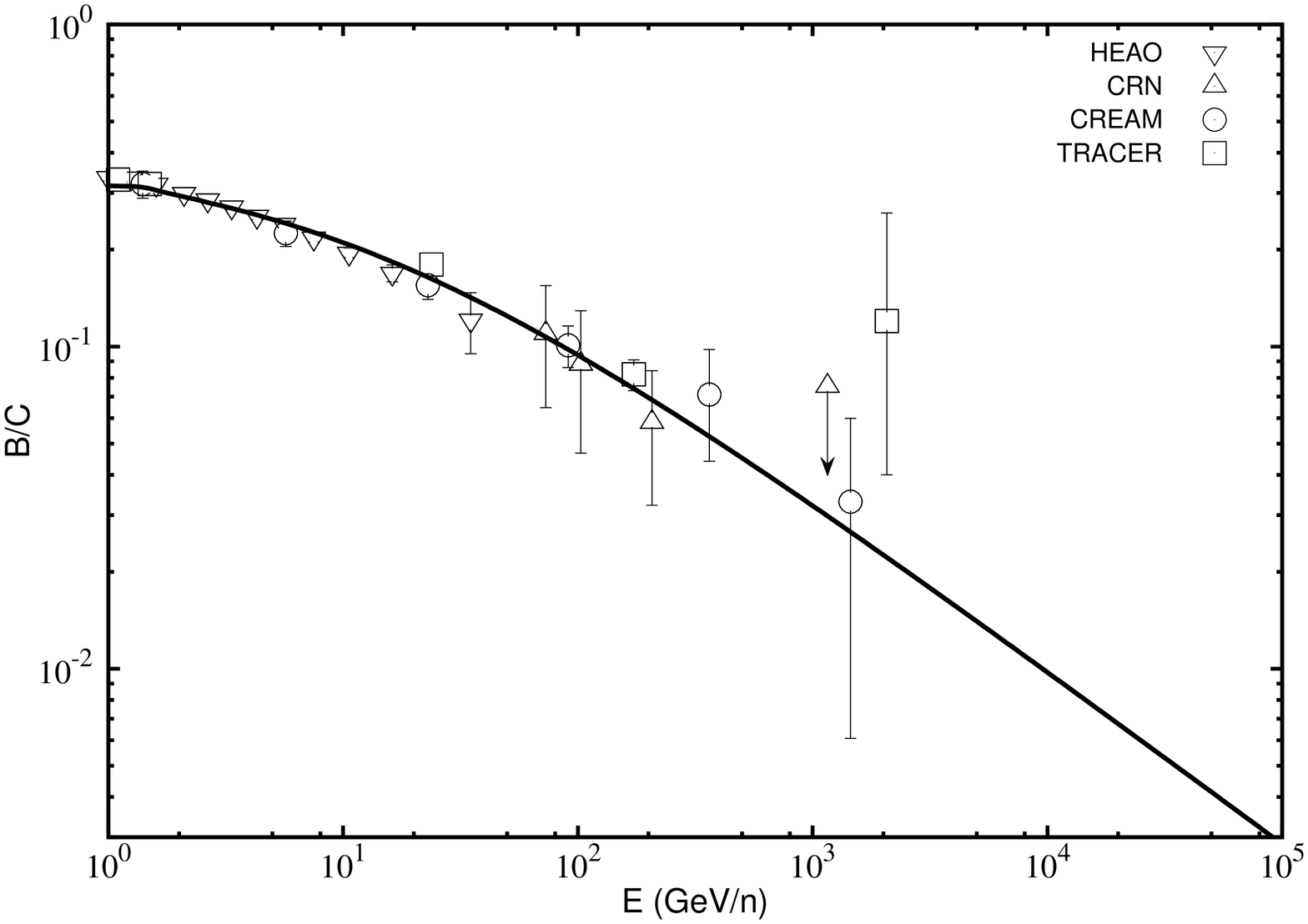}\\
\caption{\label {fig1} Boron-to-carbon ratio. Data: HEAO \citep{Engelmann1990}, CRN \citep{Swordy1990}, CREAM \citep{Ahn2008} and TRACER \citep{Obermeier2011}. {\it Thick-solid line}: Our calculation.}
\end{figure}

In our model which considers the effect of nearby sources on the observed cosmic rays, it is important to note that nearby sources can produce noticeable affects mostly on the primary spectrum. The effect on the secondaries can be neglected. This is because for cosmic rays produced by nearby sources located within  $\sim 1$ kpc from the Earth, the nuclear spallation time is much longer than the propagation time to the Earth. Therefore, the primaries do not get enough time for spallation before they reach us. By the time they undergo spallation with the interstellar matter, they have already travelled so far that the resulting secondary flux reaching the Earth is almost negligible. A detailed calculation on the resulting secondary flux can be found in \citealp{Thoudam2008}. Thus, we assume that the secondary cosmic rays that we observe are produced entirely by the background primary cosmic rays. With this assumption, the secondary spectrum can be used to determine the contribution of the background cosmic rays to the observed primary spectrum. Once the background contribution has been fixed, we can then set a limit on the contribution of nearby sources since the observed spectrum is taken to be equal to the background plus the local component. Thus, the secondary spectrum can put a constraint on the contribution of the nearby sources to the observed primary spectrum. This will be demonstrated in the following, taking boron and their primary nuclei, carbon and oxygen, as an example.
\begin{figure}
\centering
\includegraphics*[width=0.315\textwidth,angle=-90,clip]{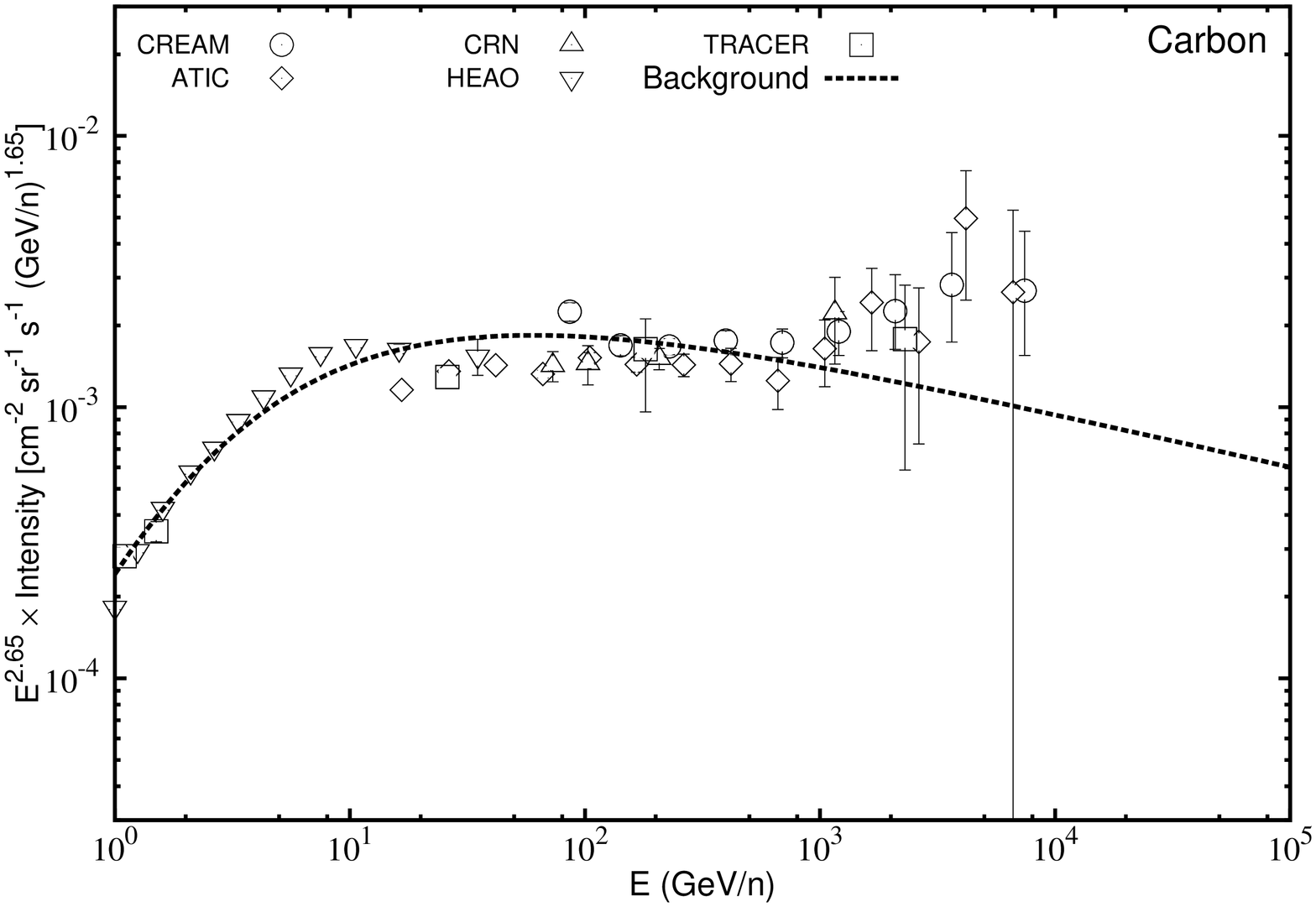}\\
\includegraphics*[width=0.315\textwidth,angle=-90,clip]{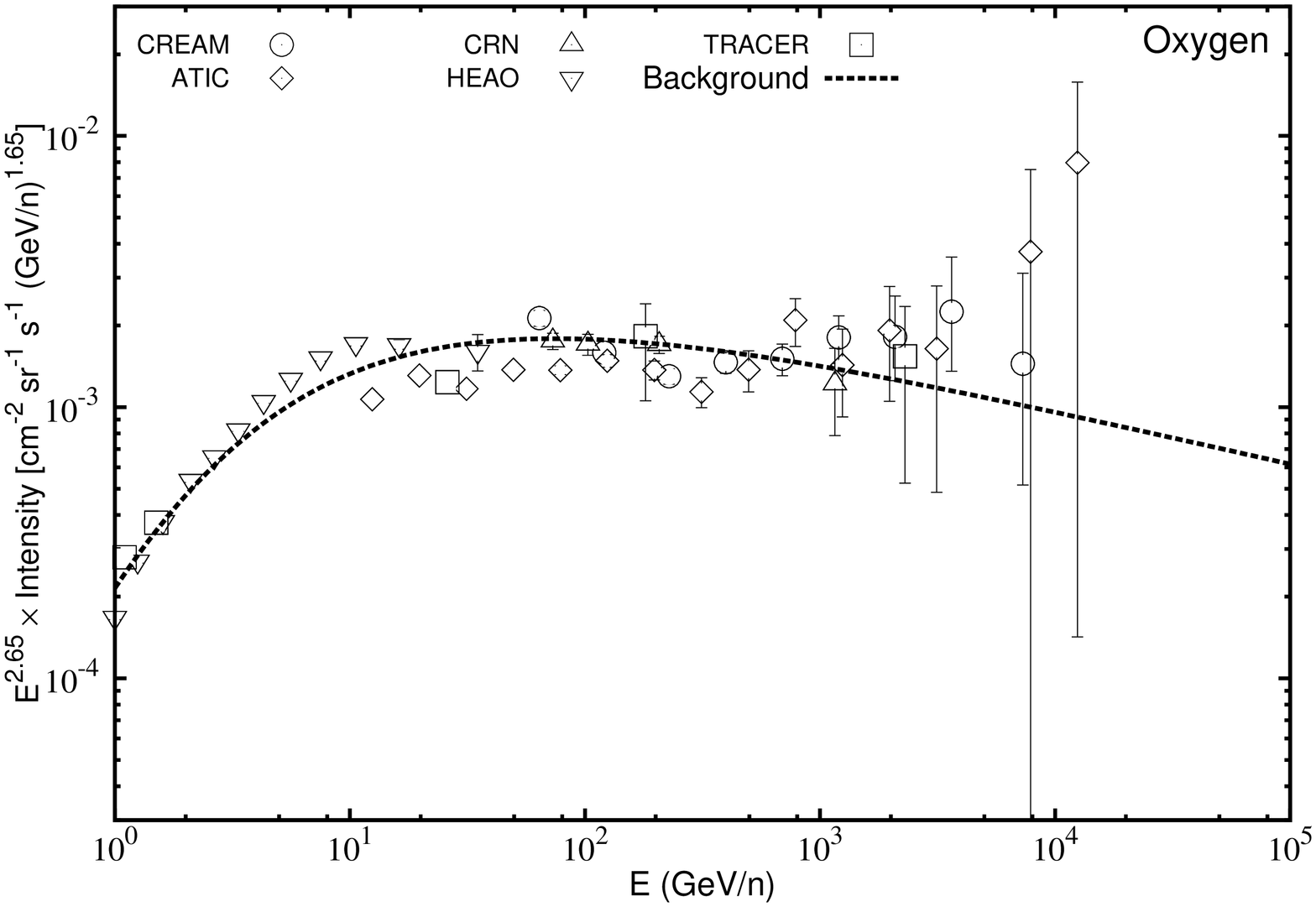}
\caption{\label {fig1} Carbon (top) and oxygen (bottom) energy spectra ($\times E^{2.65}$). Data: CREAM \citep{Ahn2009}, ATIC \citep{Panov2007}, CRN \citep{Mueller1991}, HEAO \citep{Engelmann1990} and TRACER \citep{Obermeier2011}. {\it Thick-dashed line}: Background spectrum.}
\end{figure}

Figure 2 shows the background spectra of carbon (top) and oxygen (bottom) calculated using Eq. (10). The dashed lines represent the results of our calculations, and the data are taken from CREAM \citep{Ahn2009}, ATIC \citep{Panov2007}, CRN \citep{Mueller1991}, HEAO \citep{Engelmann1990} and TRACER \citep{Obermeier2011}. The calculations assume the same source parameters for both elements. The source spectral index and the source power are chosen such that the resulting boron spectrum best explain the measured boron data up to $\sim 200$ GeV/n where the uncertainties in the measurements are small. The boron spectrum is shown in Figure 3, where the line represents our calculation and the measurements are taken from HEAO \citep{Engelmann1990}, CRN \citep{Swordy1990} and TRACER \citep{Obermeier2011}. We find that taking the source index of $\Gamma=2.31$, and the source power $Sf_{C(O)}=4.85\times 10^{48}$ ergs Myr$^{-1}$ kpc$^{-2}$ produces a good fit to the boron data up to few hundred GeV/n, where $S$ represents the surface density of the supernova explosion rate as introduced in section 4.1, $f$ is the cosmic-ray injection efficiency in units of $10^{51}$ ergs which is defined as the amount of supernova explosion energy converted into the primary species, and the subscript $C(O)$ denotes carbon (oxygen). In the present model where secondaries are assumed to be produced only in the interstellar medium, it is difficult to explain the apparent hardening of the boron spectrum in the TeV/n region indicated by the highest energy point from TRACER. This hardening  might be an effect of secondary production inside supernova remnants or re-acceleration of background secondaries by strong supernova shock waves \citep{Wandel1987, Berezhko2003}, which produces an additional hard component of secondaries.  

Although we did not normalize our calculations either on the carbon or the oxygen data, it can be seen from Figure 2 that both the background spectra are already in very nice agreement with the respective measurements up to $\sim 1$ TeV/n. This good agreement between the data and the background component has already set a very tight constraint on the contribution of  nearby sources at least below $\sim 1$ TeV/n. If we allow the maximum contribution of nearby sources to be $\sim 10\%$ of the total observed spectrum at $10$ GeV/n (see next section), the carbon or oxygen injection efficiency of supernova remnants is constraint to a value of $f_{C(O)}=0.063\%$, and the supernova surface density required to maintain the background is obtained to be  $S=7.7$ Myr$^{-1}$ kpc$^{-2}$. The latter corresponds to a supernova explosion rate of $0.97$ per century in the Galaxy.
\begin{figure}
\centering
\includegraphics*[width=0.315\textwidth,angle=-90,clip]{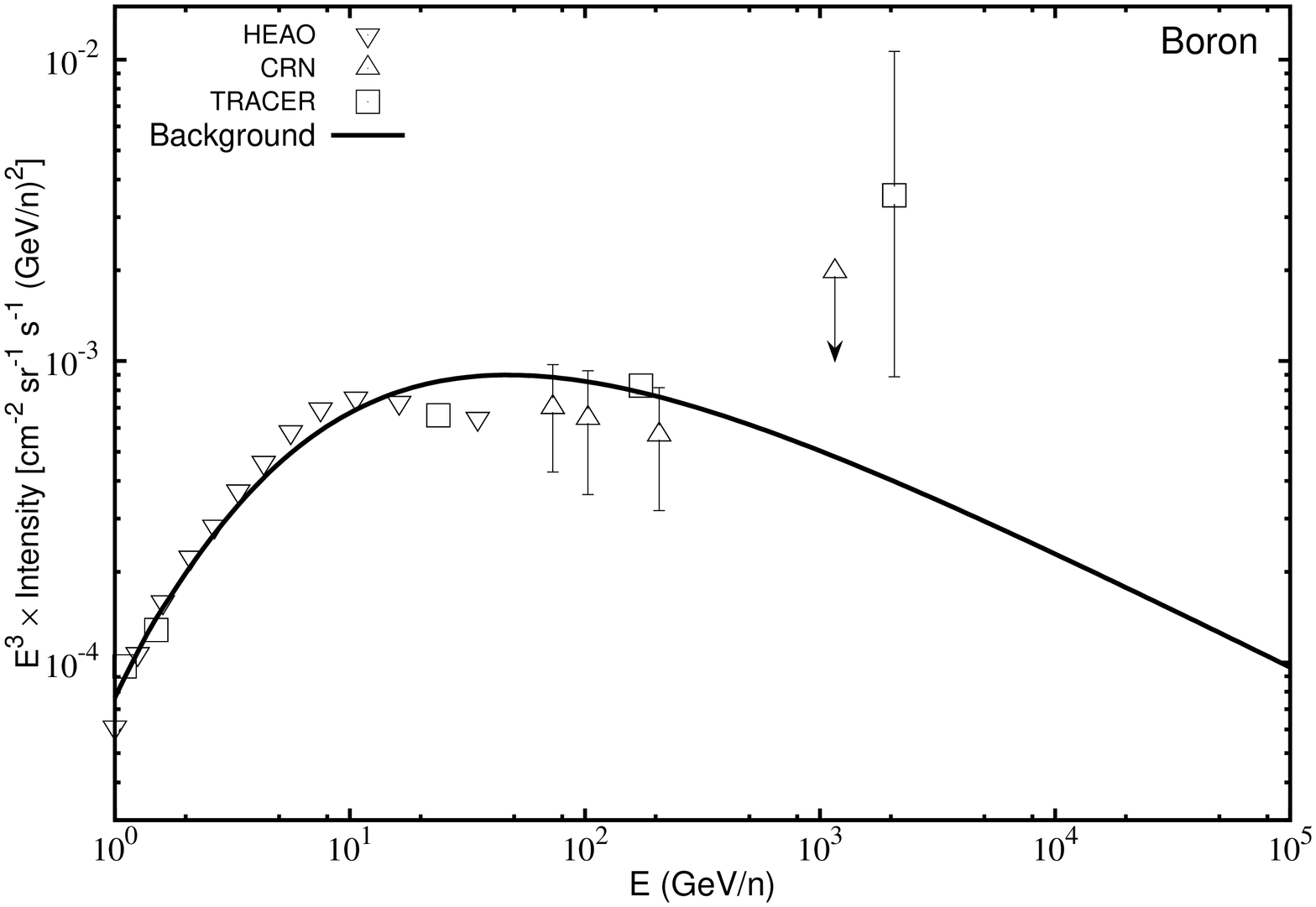}
\caption{\label {fig1} Boron energy spectrum ($\times E^{3}$). Data: HEAO \citep{Engelmann1990}, CRN \citep{Swordy1990} and TRACER \citep{Obermeier2011}. {\it Thick-solid line}: Background spectrum.}
\end{figure}

It should be noted that different nearby supernova remnants may contribute at different energies. The constraint on the cosmic-ray injection efficiency that we just derived is based on the contribution of those remnants contributing at low energies taken at  $10$ GeV/n. But, if we assume an equal injection efficiency for all the supernova remnants in the Galaxy, the same constraint can also apply to those nearby remnants contributing at higher energies, thereby also putting a limit on their contribution to the observed cosmic rays. Under this constraint, the total contribution of the nearby supernova remnants listed in Table 1 to the observed cosmic rays will be calculated. In all our calculations hereafter, the supernova rate will be taken to be the same as given above, while the injection efficiencies and the spectral indices will be allowed to vary for different cosmic-ray species and optimized based on their respective data. 

\section{Results for carbon, oxygen and iron nuclei}
In Figure 2, we can see that although the background components agree nicely with the data up to $\sim 1$ TeV/n, at higher energies there is a discrepancy between the data and the calculations. The data seem to show some excess above $\sim 1$ TeV/n. This excess or hardening in the spectrum can be explained if we include the contribution of the nearby supernova remnants as shown in Figure 4. In the figure, the thin-solid lines represent the total contribution of the nearby supernova remnants, the thick-dashed lines represent the background spectrum and the thick-solid lines represent the total nearby plus background spectrum. It can be noticed that the nearby remnants contribute mostly above $\sim 500$ GeV/n, explaining the observed spectral hardening.

In Figure 4, the thin-dashed lines show the contribution of the dominant supernova remnants at different energies. At energies below $\sim 300$ GeV/n, the nearby contribution is dominated by Loop1 and Monogem, while above, the main contributions come from Vela and G299.2-2.9. The cut-off at low energies in the case of Vela and G299.2-2.9 is largely due to energy-dependent cosmic-ray escape from the remnants with some effect of slow propagation at those energies, and the high energy fall-off is mainly because of the energy-dependent propagation effect (see section 3). 

The model parameters used in our calculation for the local component are discussed as follows. The value of $t_{sed}$ depends on the initial shock velocity of the supernova remnant, the initial ejecta mass, and the gas density of the surrounding interstellar medium. Typical values fall in the range of $\sim (100- 10^3)$ yr, and we consider $t_{sed}=500$ yr for the present work. For the initial shock velocity, we assume a value of $u_0=10^9$ cm/s. These values give the cosmic-ray escape times in the range of $t_{esc}=(500-10^5)$ yr, and $R_{esc}$ in the range of $\sim(5-100)$ pc. The cosmic-ray escape parameter $\alpha$ is kept as a free parameter, but its value is assumed to be the same for all the cosmic-ray species. For a given cosmic-ray species, we take the same injection efficiency and the same source index as those used in the calculation for the background component. And, all the nearby supernova remnants considered in our study are assumed to have the same set of model parameters.
\begin{figure}
\centering
\includegraphics*[width=0.315\textwidth,angle=-90,clip]{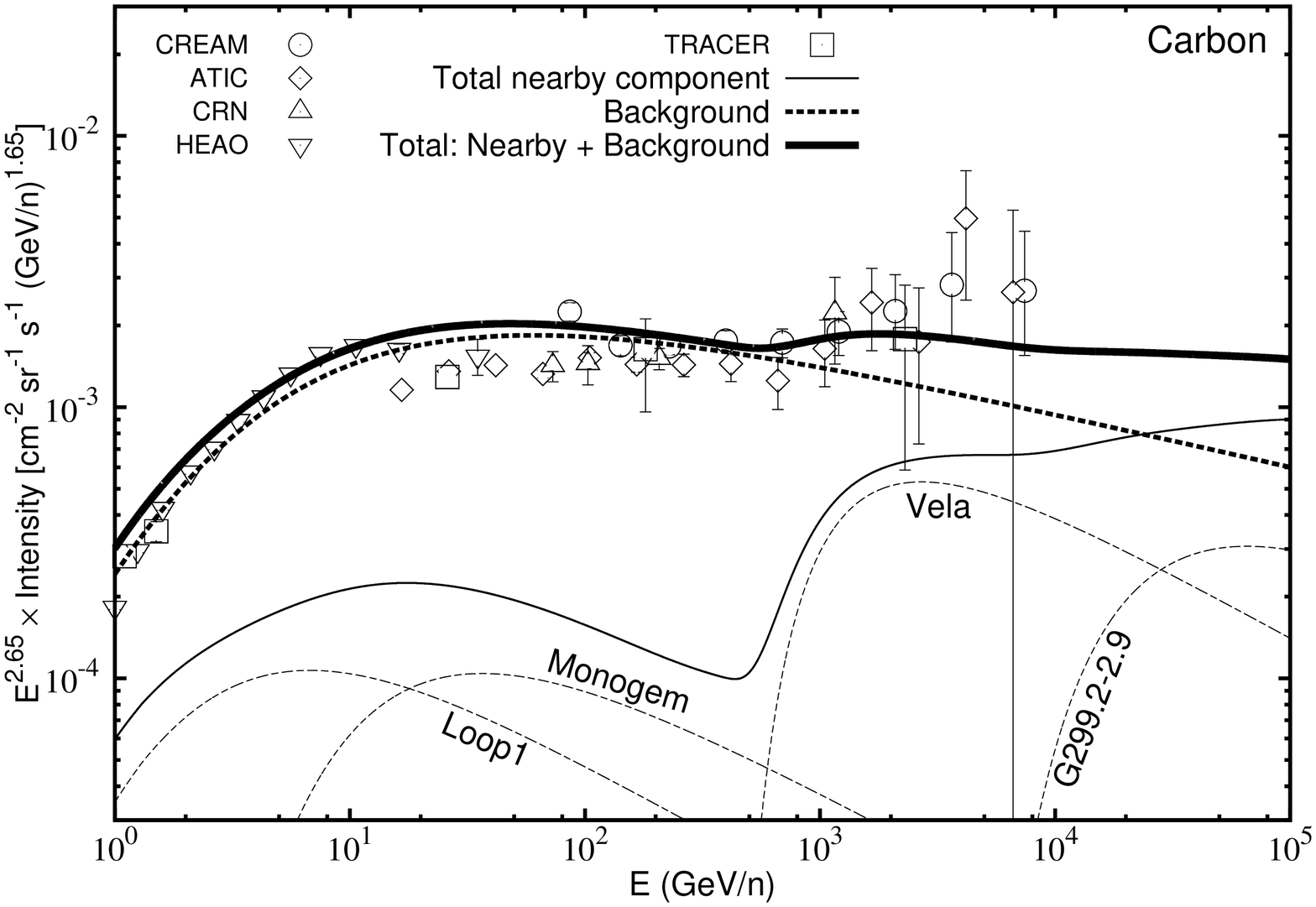}\\
\includegraphics*[width=0.315\textwidth,angle=-90,clip]{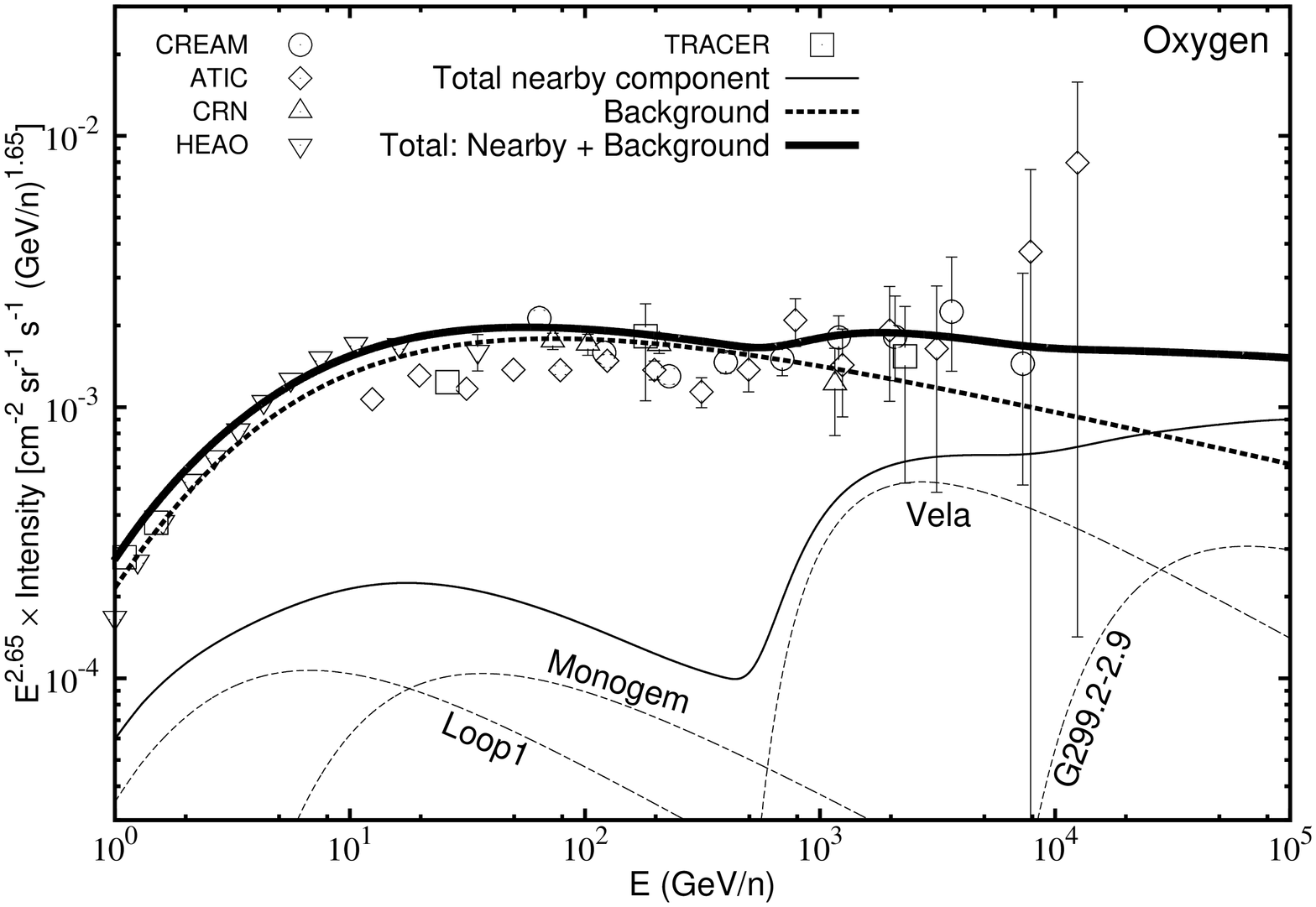}
\caption{\label {fig1} Carbon (top) and oxygen (bottom) energy spectra ($\times E^{2.65}$). {\it Thick-dashed line}: Background spectrum. {\it Thin-solid line}: Total nearby component. {\it Thick-solid line}:  Total nearby plus background. {\it Thin-dashed lines}: Dominant nearby supernova remnants.}
\end{figure}

For the results shown in Figure 4, $\Gamma_{C(O)}=2.31$, $f_{C(O)}=0.063\%$, $S=7.7$ Myr$^{-1}$ kpc$^{-2}$ and $\alpha=2.4$. This value of $\alpha$ gives particle escape times of  $t_{esc}=(500-10^5)$ yr for particles of energies $0.5$ PeV/n to $1.5$ GeV/n. The present value of $\alpha$ is slightly larger than the value of $2.2$ adopted in \citetalias{Thoudam2012a}. A larger value is required due to the smaller value for the diffusion coefficient used in the present study. Physically speaking, larger $\alpha$ means shorter confinement within the remnant, while smaller $D$ means longer time for cosmic rays to reach the Earth. Therefore, in order to explain the spectral hardening above $\sim 500$ GeV/n, which in our model is due to the contribution of the nearby sources, a larger $\alpha$ is required to compensate the effect of a smaller $D$. This will become more clear in the next section, when we compare our present results for protons and helium nuclei with those obtained in \citetalias{Thoudam2012a}.
\begin{figure}
\centering
\includegraphics*[width=0.315\textwidth,angle=-90,clip]{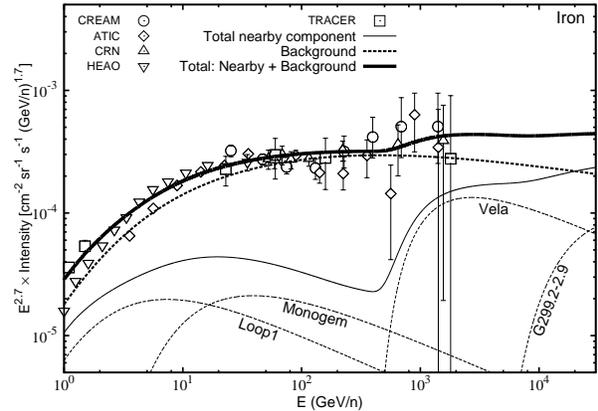}
\caption{\label {fig1} Iron energy spectrum ($\times E^{2.7}$). Data: CREAM \citep{Ahn2009}, ATIC \citep{Panov2007}, CRN \citep{Mueller1991}, HEAO \citep{Engelmann1990} and TRACER \citep{Obermeier2011}. {\it Thick-dashed line}: Background spectrum. {\it Thin-solid line}: Total nearby component. {\it Thick-solid line}:  Total nearby plus background. {\it Thin-dashed lines}: Dominant nearby supernova remnants.}
\end{figure}

For iron, the result is shown in Figure 5. We take the same source index of $\Gamma_{Fe}=2.31$ as in the case of carbon and oxygen, and an injection efficiency of $f_{Fe}=0.011\%$. All other model parameters remain the same as in Figure 4. Also, in the case of iron, we can see that the nearby supernova remnants produce a spectral hardening above $\sim 500$ GeV/n. Future sensitive measurements at high energies can provide a crucial check of our prediction.

\section{Results for proton and helium nuclei}
In Figure 6, we show our results for protons (top) and helium nuclei (bottom). The thick-dashed and the thin-solid lines  represent the background and the total nearby contributions respectively, and the thick-solid line shows the total sum of the background and the nearby components. The contributions from the nearby dominant sources are also shown by the thin-dashed lines. The measurements data are from CREAM \citep{Yoon2011}, ATIC\footnote{Data taken from the compilation by \citealp{Strong2009}.} \citep{Panov2007}, PAMELA\footnotemark[\value{footnote}] \citep{Adriani2011}, and AMS \citep{Alcaraz2000, Aguilar2002}. The results look very similar to those obtained for the heavier nuclei in section 7. Except for the source index and the injection efficiency, all other model parameters are taken to be the same as for the heavier nuclei. For protons, we find that taking $\Gamma_P=2.27$ and $f_P=17.5\%$ produces a good fit to the data, while for helium the best fit parameters are found to be $\Gamma_{He}=2.21$ and $f_{He}=1.75\%$. For comparison, these values are listed in Table 2 along with those obtained for carbon, oxygen and iron.
\begin{figure}
\centering
\includegraphics*[width=0.315\textwidth,angle=-90,clip]{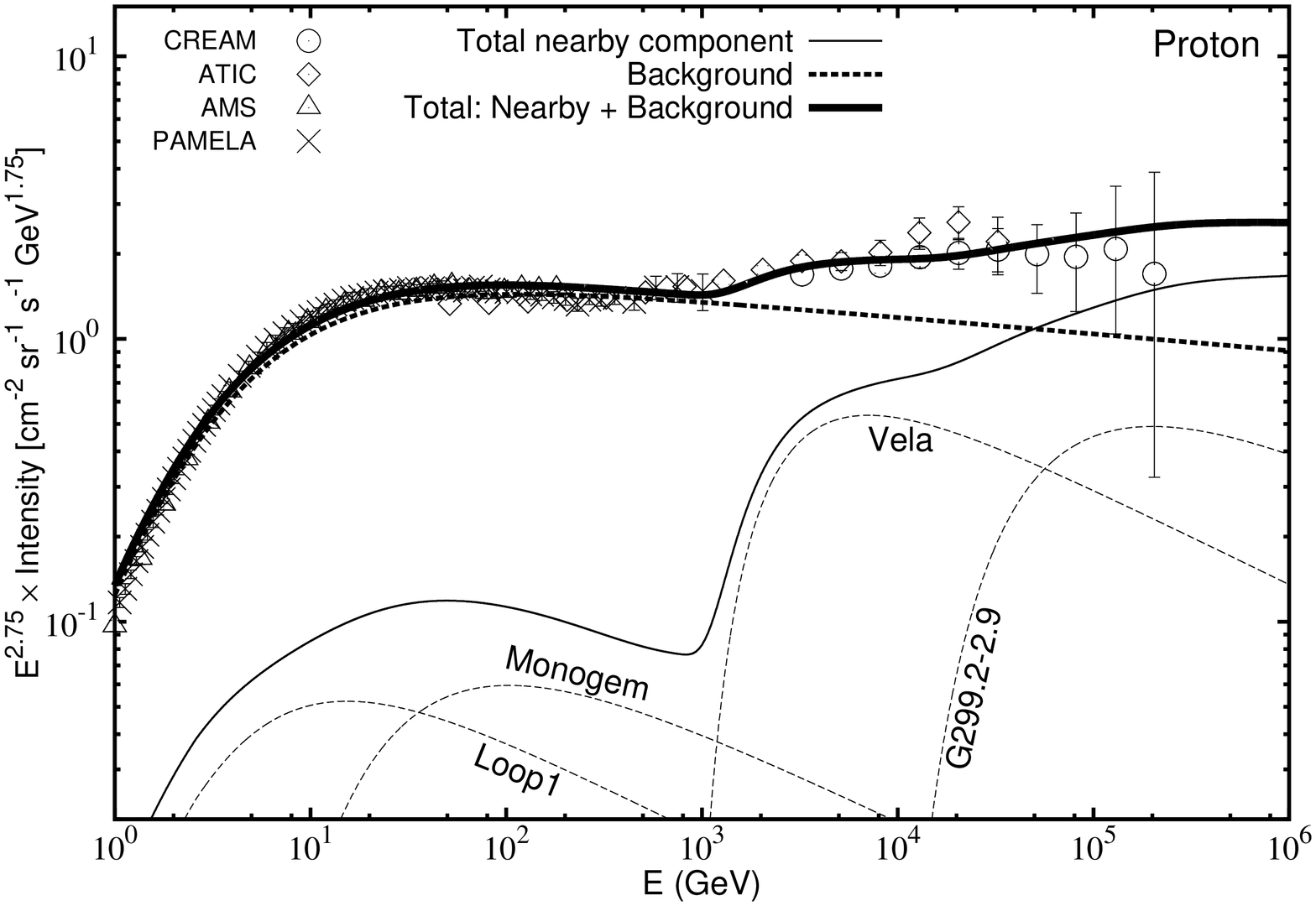}\\
\includegraphics*[width=0.315\textwidth,angle=-90,clip]{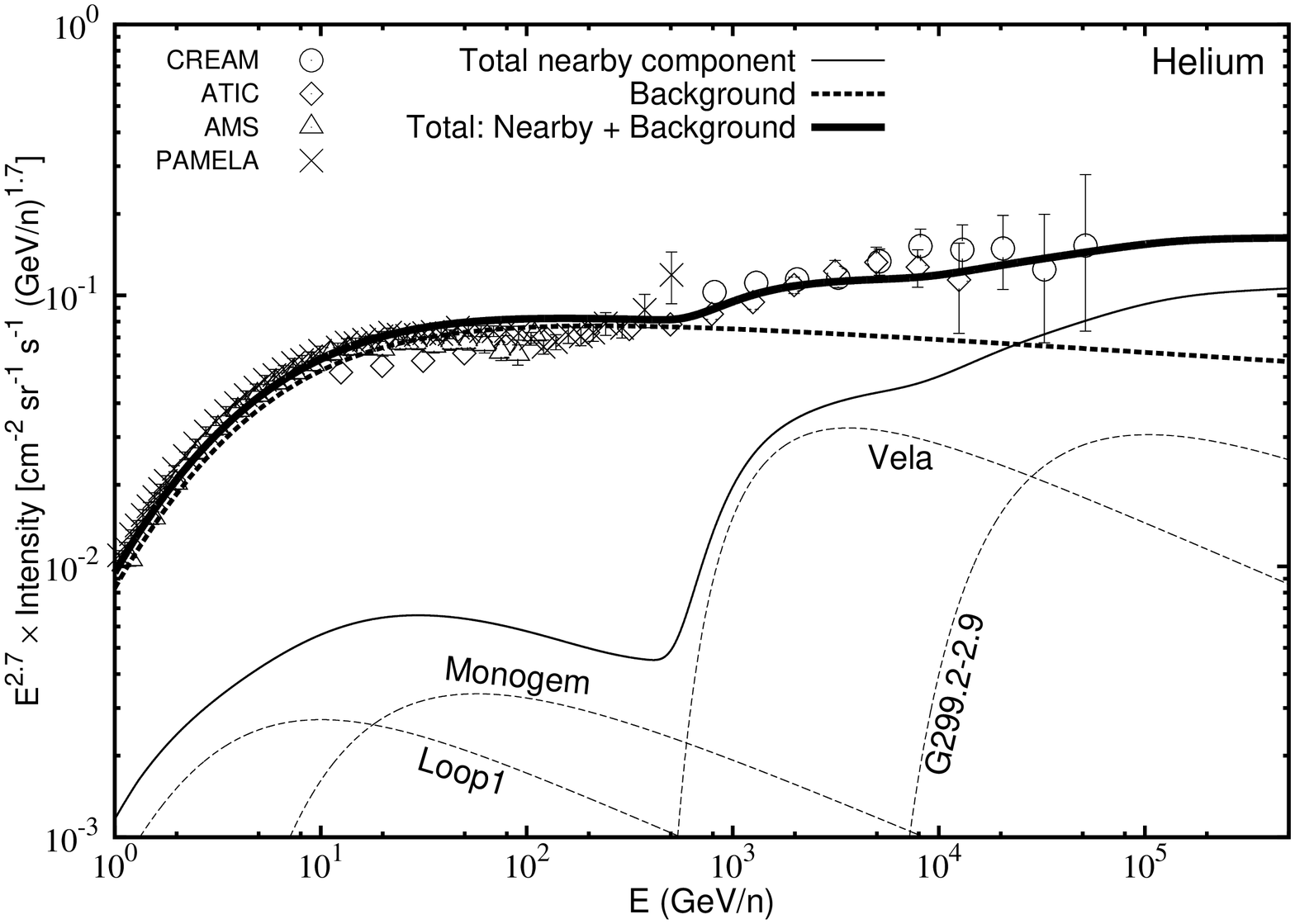}
\caption{\label {fig1} Proton ($\times E^{2.75}$, top) and helium ($\times E^{2.7}$, bottom) energy spectra. Data: CREAM \citep{Yoon2011}, ATIC \citep{Panov2007}, PAMELA \citep{Adriani2011}, and AMS \citep{Alcaraz2000, Aguilar2002}. {\it Thick-dashed line}: Background spectrum. {\it Thin-solid line}: Total nearby component. {\it Thick-solid line}:  Total nearby plus background. {\it Thin-dashed lines}: Dominant nearby supernova remnants.}
\end{figure}

In general, our present results are quite similar to the results presented in \citetalias{Thoudam2012a}. For instance, the nearby supernova remnants show dominant contribution at energies $\gtrsim (0.5-1)$ TeV/n, thereby explaining the observed spectral hardening. Moreover, the helium spectrum shows a hardening at lower energy/nucleon with respect to the proton spectrum which, in our model, is attributed mainly to the early escape times of helium nuclei from the supernova remnants relative to the protons. It might be recalled from \citetalias{Thoudam2012a}, and also discussed in section 3 of this paper that in our model, such a spectral hardening at lower energies/nucleon is expected for all heavier primaries $(A>1)$ whose escape times are shorter than the time for protons by a factor of $(A/Z)^{-1/\alpha}$ at the same energy/nucleon. Also, both our present and the previous studies show that the main contribution at high energies comes from the Vela and G299.2-2.9 remnants, and that the spectral hardening does not continue up to very high energies. Available measurements also seem to support these results.
\begin{table}
\centering
\caption{Source spectral indices $\Gamma$ and injection efficiencies $f$ for the various cosmic-ray nuclei  considered in our study.}
\bigskip
\begin{tabular}{c|c|c}
\hline
Nuclei	&	$\Gamma$		&	$f$ ($\times 10^{49}$ ergs) \\         
\hline
Proton		&	$2.27$	&	$17.5$\\
Helium		&	$2.21$	&	$1.75$\\
Carbon		&	$2.31$	&	$0.063$\\
Oxygen		&	$2.31$	&	$0.063$\\
Iron		&	$2.31$	&	$0.011$\\
\hline
\end{tabular}
\end{table}

However, some basic differences can be noticed. First, the background spectrum is steeper in the present case. At high energies, the proton background follows a spectral index of $2.82$ and the helium background has an index of $2.75$. To be compared, the background indices were obtained as $2.75$ and $2.68$ respectively in  \citetalias{Thoudam2012a}. It may be recalled that in \citetalias{Thoudam2012a}, the backgrounds were obtained by fitting the measured spectra between $(20-200)$ GeV/n and their consistency with the low energy data below $\sim 20$ GeV/n was not checked. From Figure 6, it can be noticed that the background adopted in the present study agrees nicely (in fact, even better when the small local component has been added) with the data down to $1$ GeV/n.

Second, in \citetalias{Thoudam2012a}, the source index for the nearby sources were obtained as $(\Gamma_b-\delta)$, where $\Gamma_b$ is the background index and $\delta$ is the diffusion index. So, for the proton background of $\Gamma_b=2.75$ and $\delta=0.6$ adopted in \citetalias{Thoudam2012a}, the source index was found to be $2.15$. This is flatter than the proton source index of $2.27$ adopted in the present study. Adopting a steeper source index suppresses the contribution from nearby sources at high energies. For the same amount of total energy injected into protons, a source with an index of $2.27$ produces $\sim 1.8$ times less number of source particles at $\sim 10$ TeV than a source with an index of $2.15$. A similar difference is also expected in the case of  helium. On the other hand, taking a smaller value of $D$ enhances the flux from a nearby source. This is clear from the discussion on Eq. (8), given in section 3 which showed that at very high energies, the particle spectrum depends on $D$ as $N\propto D^{-3/2}$. After detailed investigation, we find that the increase in the particle flux in the present study due to smaller $D$ is almost equal to the decrease in the flux due to the steeper source index. Because of these two almost equal and opposite effects, we can still explain the spectral hardening of helium with an injection efficiency very close to that used in \citetalias{Thoudam2012a}. For protons, we need an injection efficiency of $17.5\%$ in the present study, which is approximately twice the value obtained in \citetalias{Thoudam2012a}. This difference is because of the lower proton background in the present study at energies above $\sim 1$ TeV as compared to the background in \citetalias{Thoudam2012a}. For helium, the background does not differ too much at TeV energies in the two studies.

Another difference results from the difference in the cosmic-ray escape parameter and the diffusion index. A smaller $\alpha$ produces a sharper low-energy cut-off and a larger $\delta$ leads to a steeper fall-off in the high-energy spectrum from a nearby source. In \citetalias{Thoudam2012a}, where we took $\alpha=2.2$ and $\delta=0.6$, this led to sharper peaks in the individual contributions of the nearby remnants, thereby resulted into stronger structures in the resultant total spectrum. In the present study, the slightly larger value of $\alpha=2.4$ and the smaller value of $\delta=0.54$ produce broader peaks in the individual contributions leading to weaker structures in the overall total spectrum.      

\section{Discussion}
We have shown that the spectral hardening of cosmic rays at TeV energies recently observed by the ATIC, CREAM, and PAMELA experiments can be due to nearby supernova remnants. Considering that cosmic rays escape from supernova remnants in an energy-dependent manner, we also show that heavier elements should exhibit spectral hardening at relatively lower energies/nucleon with respect to protons, and that the hardening might not continue  up to very high energies. These results also seem to agree with the measured data.

In general, the results obtained in this paper agree very well with those presented in \citetalias{Thoudam2012a}. Our present study involves a detailed calculation of the background cosmic rays unlike in \citetalias{Thoudam2012a}, and also follow a consistent treatment of the cosmic-ray source spectrum for the background and the nearby sources. Our results are found to be consistent with the observed data over a wide range in energy from $1$ GeV/n to $\sim 10^5$ GeV/n for a reasonable set of model parameters. Our calculation requires a supernova explosion rate of $\sim 1$ per century in the Galaxy, and cosmic-ray injection efficiencies of $f_P=17.5\%$ for protons, $f_{He}=1.75\%$ for helium nuclei which is exactly $10\%$ of the proton value, $f_{C(O)}=0.063\%$ for carbon and oxygen, and $f_{Fe}=0.011\%$ for iron. The required source index for protons is $\Gamma_P=2.27$ and for helium nuclei, $\Gamma_{He}=2.21$. For carbon, oxygen and iron, we determined the same source index of $2.31$. The required source indices of $\sim (2.2-2.3)$ in the present study are slightly steeper than a value of $\Gamma=2.0-2.2$ predicted by DSA theory. Actually, even a larger value of $\Gamma\sim (2.4-2.5)$ is favored by the high level of cosmic-ray isotropy observed between around $1$ and $100$ TeV which in turn suggests a smaller value of the diffusion index of $\delta\sim$ $0.3$ to $0.4$ \citep{Ptuskin2006}. This discrepancy between observation and theory is still not clearly understood.

Our model predictions are expected to be different in many respects from those of other existing models. Models which are based either on the hardening in the source spectrum or changes in the diffusion properties of cosmic rays at high energies will produce a spectrum that remains hard up to very high energies (\citealp{Ohira2011}, \citealp{Yuan2011}, \citealp{Ave2009}). But, although not very significant, the CREAM data seems to indicate that the spectral hardening does not continue beyond few tens of TeV/n, which in general is in good agreement with our prediction. In Figure 7, we compare our result for proton spectrum (double-dashed line) with the predictions of other models: Model I (thick-dashed line) and Model II (thin-solid line). Model I represents models with a hardened source spectrum above a certain energy, but assumes the same diffusion coefficient as in our model. Model II represents models which incorporate a break or hardening in the diffusion coefficient. The source spectrum in Model II is taken to be the same as in our model. Also shown in Figure 7 for reference is the background spectrum obtained in our model. To reproduce the same measured spectrum, Models I and II are chosen to have their respective breaks in the source spectrum or in the diffusion coefficient at the same energy $E_b=850$ GeV/n. It can be noticed that our result shows significant difference mostly at energies above $\sim 0.1$ PeV. Thus, if the cosmic-ray spectrum exhibit an exponential cut-off below $\sim 0.1$ PeV, our result will not be significantly different from the others. However, detailed studies on the origin of the cosmic-ray knee suggest a cut-off at energies around $3-5$ PeV \citep{Hoerandel2003}.
\begin{figure}
\centering
\includegraphics*[width=0.315\textwidth,angle=-90,clip]{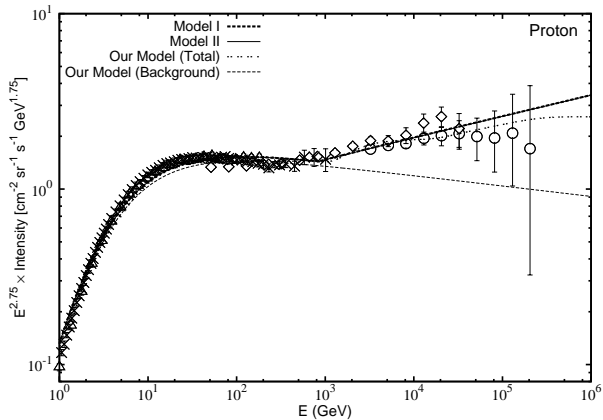}
\caption{\label {fig1} Proton energy spectrum under different models. {\it Thick-dashed line}: Model I. {\it Thin-solid line}: Model II. {\it Double-dashed line}: Total background plus nearby spectrum in our model. {\it Thin-dashed line}: Our background spectrum. Data as given in Figure 6 (top).}
\end{figure}

Even more different between the different models will be the secondary-to-primary ratios and the secondary spectra. In the standard model of cosmic-ray propagation, the secondary-to-primary ratio is independent of the source parameters and gives  a good measure of the cosmic-ray diffusion coefficient in the Galaxy. However, in our model, which considers the effect of the nearby sources, the ratio may deviate from the standard result. This is because the nearby sources can affect only the primary spectrum, and their effect on the secondaries is negligible. Thus, we expect to see a steepening in the ratio in the energy region where the nearby contribution on the primary spectrum is significant. This is shown in Figure 8 for the boron-to-carbon ratio. Notice the  significant steepening in our model above $\sim 500$ GeV/n with respect to Model I even though both the models assume the same diffusion coefficient. The result for Model II is even more flatter at high energies, reflecting  the harder value of diffusion coefficient above $E_b=850$ GeV/n assumed in the model.
\begin{figure}
\centering
\includegraphics*[width=0.315\textwidth,angle=-90,clip]{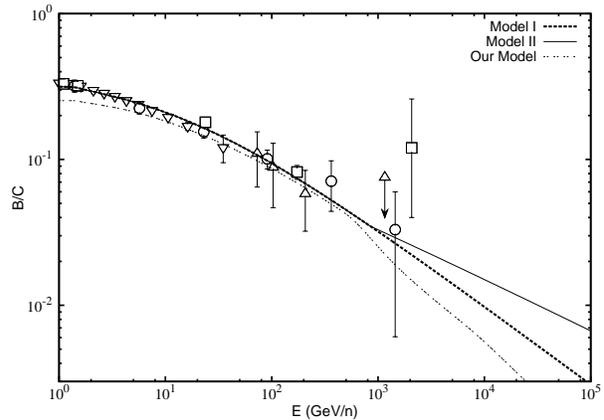}
\caption{\label {fig1} Boron-to-carbon ratio under different models. {\it Thick-dashed line}: Model I. {\it Thin-solid line}: Model II. {\it Double-dashed line}: Our model. Data as given in Figure 1.}
\end{figure}
\begin{figure}
\centering
\includegraphics*[width=0.315\textwidth,angle=-90,clip]{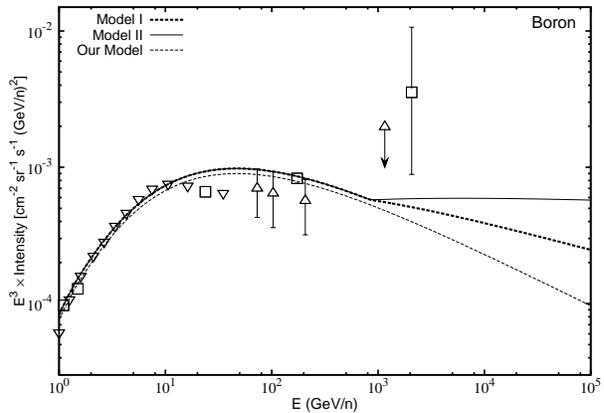}
\caption{\label {fig1} Boron energy spectrum under different models. {\it Thick-dashed line}: Model I. {\it Thin-solid line}: Model II. {\it Double-dashed line}: Our model. Data as given in Figure 3.}
\end{figure}

For an equilibrium primary spectrum $N_p\propto E^{-\gamma}$, and a diffusion coefficient $D\propto E^\delta$, the secondary spectrum in the Galaxy follows $N_s\propto E^{-(\gamma+\delta)}$. This shows that for a fixed diffusion coefficient, the shape of the secondary spectrum is determined by the shape of the primary spectrum. Therefore, models that consider the same diffusion coefficient, but different $N_p$ will produce different $N_s$. Since the background primary spectrum above $E_b$ in our model is steeper than in Model I (see Figure 7, thin dashed and thick-dashed lines), the secondary spectrum is expected to be steeper above $E_b$ in our model. This is shown in Figure 9 for the boron spectrum, where the thick-dashed line represents Model I, and the thin-dashed line represents our model. The difference is expected to be even more significant if we compare with Model II, which assumes a harder diffusion index above $E_b$. The thin-solid line in Figure 9 represents Model II. Similar differences are also expected in other types of secondary nuclear species such as sub-iron, and anti-protons as shown in \citealp{Vladimirov2012}, which also discussed the effects of different models on various observed properties of cosmic rays including the secondary-to-primary ratios and the diffuse gamma-ray emissions. 
\begin{figure}
\centering
\includegraphics*[width=0.315\textwidth,angle=-90,clip]{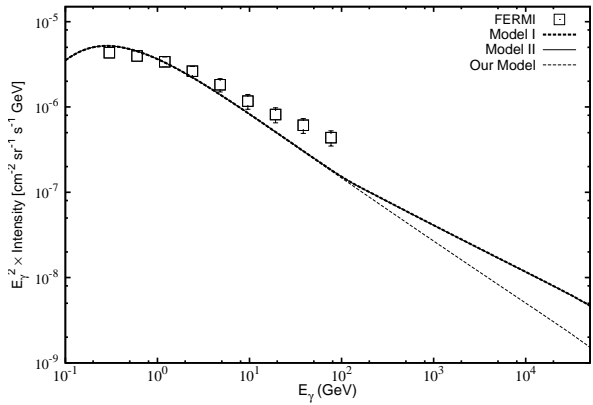}
\caption{\label {fig1} Galactic diffuse gamma-ray spectrum under different models. {\it Thick-dashed line}: Model I. {\it Thin-solid line}: Model II. {\it Double-dashed line}: Our model. The data represents the gamma-ray intensity measured by FERMI for Galactic latitudes $|b|\geq 10^\circ$ \citep{Abdo2010}.}
\end{figure}

The Galactic diffuse gamma-ray spectrum in our model is also expected to be different from those calculated using other models. If the diffuse emission is dominated by gamma-rays produced from the decay of $\pi^0$ mesons, then their spectrum at high energies would largely follow that of the primary protons. As the background spectrum in our model is steeper above $E_b$ than the spectrum adopted in other models, our diffuse gamma-ray spectrum will be steeper above $E_b$. This is shown in Figure 10 where the thin-dashed line represents our model, and the thick dashed and thin-solid lines represent Models I and II respectively. The data represents the gamma-ray intensity for Galactic latitudes $|b|\geq 10^\circ$ measured by the FERMI experiment \citep{Abdo2010}. All results plotted in Figure 10 are normalized to the data at $1.2$ GeV. The calculations use the different proton spectra shown in Figure 7, and the gamma-ray production cross-section is taken from Kelner et al. 2006. In Figure 10, we can see that Models I and II give almost the same result, and produce a harder gamma-ray spectrum with respect to our model above $\sim 100$ GeV. This difference can be checked by future measurements at high energies, and can distinguish our model from others. The data in Figure 10 show some excess above the model predictions between $\sim (10-100)$ GeV. Although it is not the aim of this paper to perform a detailed modeling of the FERMI data, it can be mentioned that the excess is most likely due to additional contributions from other processes such as bremsstrahlung, inverse compton and unresolved point sources, which are neglected in our calculations. Detailed calculations involving all the possible contributions have shown that the diffuse gamma-rays up to $\sim 100$ GeV measured by FERMI from different regions in the Galaxy can be explained using a single power-law cosmic-ray spectrum without any break above a few GeV \citep[see also \citealp{Vladimirov2012}]{Ackermann2012}.

\section{Conclusion}
We have presented a detailed and improved version of our previous work presented in \citetalias{Thoudam2012a} \citep{Thoudam2012a} where we showed that the spectral hardening of cosmic rays observed at TeV energies can be a local effect due to nearby supernova remnants. Unlike in \citetalias{Thoudam2012a} where the cosmic-ray background was obtained by merely fitting the low-energy data, the present work involves a detailed calculation of both, the background and the local components, considering consistent cosmic-ray source parameters between the two components.

In addition to the results for protons and helium nuclei (which were also studied in \citetalias{Thoudam2012a}), we have also presented results for heavier cosmic-ray species, such as boron, carbon, oxygen, and iron nuclei. Unlike in other existing models, we have shown that heavier nuclei should exhibit hardening at a lower energy/nucleon as compared to protons, and also that the hardening may not continue up to very high energies for all the species. Although not very significant, the available data seem to suggest our findings. Moreover, we have also shown that our results on the secondary-to-primary ratios, the secondary spectra, and the Galactic diffuse gamma-ray spectrum at high energies are expected to be different from the predictions of other models. Future sensitive high-energy measurements on these quantities can differentiate our model from others.

\end{document}